\documentclass[12pt]{article}
\usepackage{amsfonts}
\usepackage{amsmath}
\usepackage{a4wide}

\input{tcilatex}

\begin{document}

\title{Reply to "Comment on 'Separability of quantum states and the
violation of Bell-type inequalities'"}
\author{Elena R. Loubenets \\
Applied Mathematics Department, \\
Moscow State Institute of Electronics and Mathematics}
\maketitle

\begin{abstract}
It is stated in [1] that the definition of "classicality" used in [2] is
"much narrower than Bell's concept of local hidden variables" and that, in
the separable quantum case, the validity of the perfect correlation form of
the original Bell inequality is necessarily linked with "the assumption of
perfect correlations if the same (quantum) observable is measured on both
sides". Here, I prove that these and other statements in [1] are misleading.
\end{abstract}

Let me first specify a misleading character of the statements in [1] on the
description of classical joint measurements in appendix of [2].

In the general formalism for the description of ideal measurements on a
physical system, both types of ideal measurements, quantum and classical,
are described in a unique manner in terms of states and observables, quantum
or classical. Every classical joint measurement is described by two
classical observables where each observable represents a classical system
property observed under the corresponding marginal experiment. In the
inequality (\textrm{A5}) in [2], the symbols $A,$ $D_{1}$ and $D_{2}$ stand
for classical observables while $\pi $ - for a classical state. Whenever a
classical joint measurement is formulated in terms of classical observables,
it does not already matter where (or on which "side") each of marginal
experiments is performed - the notion of a classical observable includes all
specifications on the corresponding marginal measurement. In [2], the
relations (\textrm{A3}) do not represent my "postulates" and do not
correspond to "the assumption of perfect correlations", as claimed in [1],
but follow from the generally accepted formula (\textrm{A1}) for the
expectation value of the product of outcomes observed under a classical
joint measurement. Under a classical joint measurement "$A$ and $D"$ of two
classical observables $A$, $D,$ the joint probability that the corresponding
real-valued outcomes $\lambda _{1}$ and $\lambda _{2}$ belong, respectively,
to subsets $B_{1},B_{2}\subseteq \mathbb{R}$ is given by the formula%
\footnote{%
Here, I use the notation in appendix of [2] and $f^{-1}(B)$ denotes the
subset $\{\theta \in \Theta :f(\theta )\in B\}$.} $\pi
(f_{A}^{-1}(B_{1})\cap f_{D}^{-1}(B_{2}))$. Therefore, under a classical
Alice/Bob joint measurement of the same classical observable $A$ on both
sides, the joint probability distribution has the form $\pi
(f_{A}^{-1}(B_{1}\cap B_{2}))$ and, in the discrete case, the outcomes on
both sides are always perfectly correlated.

The only difference between the classical joint measurement situations,
discussed in appendix of [2] and in J. Bell's presentation [3], constitutes
the number of classical observables describing these measurement situations.
Namely, the classical joint measurement situation discussed in appendix of
[2] is described by three classical observables $A,$ $D_{1}$ and $D_{2}$
while the classical joint measurement situation in [3] - by four classical
observables represented on a variable set $\Theta $ by functions $%
f_{1}(\theta ,a),$ $f_{1}(\theta ,b),$ $f_{2}(\theta ,b),$ $f_{2}(\theta ,c)$
(in the notation of [3]).

Consider now the reasoning of [2] on, in general, "non-classical" character
of behaviour demonstrated by a separable quantum state under quantum
Alice/Bob measurement situations.

As it is proved in [2] in a general setting\footnote{%
The original Bell proof corresponds only to the case of dichotomic classical
observables admitting values $\pm \lambda .$}, under any ideal classical
joint measurement situation described by three classical observables $A$, $%
D_{1}$, $D_{2}$ the product expectation values satisfy the inequality (%
\textrm{A6}) in [2] for every classical state $\pi .$ The classical
probabilistic constraint (\textrm{A6}) in [2] is, in particular, true under
any three classical Alice/Bob joint measurements $(A,D_{1}),$ $(A,$ $D_{2})$
and $(D_{1},D_{2})$ where Alice and Bob measure observables standing in the
first and the second places of a pair, respectively.

Let Alice and Bob perform three quantum joint measurements $(A,D_{1}),$ $(A,$
$D_{2})$ and $(D_{1},D_{2})$ on identical quantum sub-systems in a quantum
state $\rho $. If, for some quantum observables $A$, $D_{1},$ $D_{2},$ the
corresponding product expectation values in a state $\rho $ violate the
classical probabilistic constraint (\textrm{A6}) in [2] then, in view of the
unified probabilistic description of quantum and classical Alice/Bob joint
measurements in terms of expectation values, states and observables, this
bipartite quantum state $\rho $ cannot be argued to exhibit "classical-like"
character of behaviour under ideal quantum Alice/Bob joint measurement
situations. The latter is equivalently true for separable and nonseparable $%
\rho .$

The statement in [2] on, in general, "non-classical" character of behaviour
of a separable state is based on the violation of Eq. (\textrm{A6}) in [2]
for the separable state (6) in [2] under three ideal Alice/Bob joint spin
measurements specified in section 2 of [2]\footnote{%
In section 2, under the joint spin measurement, specified by parameters $%
\vartheta _{a}$ and $\vartheta _{b}$ on the sides of Alice and Bob,
respectively, Alice measures the quantum observable $\{J^{(\vartheta
_{a})}\otimes I\}_{sym}$ while Bob - $\{J^{(\vartheta _{b})}\otimes
I\}_{sym} $ (cf. the notation and discussion in [2]).}.

It is, however, argued in [1] that the violation of Eq. (\textrm{A6}) in [2]
for a separable quantum state "does not demonstrate 'non-classical'
behaviour in any usual sense of the word". In [1], this conclusion is built
up on the existence, for quantum Alice/Bob joint measurements on a separable
state, of a Bell local hidden variable (\textit{LHV}) model formulated, in
general, in [3].

Recall that, in the frame of a Bell \textit{LHV} model, one and the same
quantum observable measured by Alice and Bob is contextually represented by
two classical observables. That is why, any ideal quantum Alice/Bob joint
measurement situation described by three quantum observables is simulated by
a classical Alice/Bob measurement situation described by four classical
observables. According to [1], "only if one has additional information on
the system under consideration", the number of simulating classical
observables can be reduced. It is then concluded in [1] that, in the
separable quantum case, the validity of Eq. (\textrm{A6}) in [2] is
necessarily linked with "the assumption of perfect correlations if the same
(quantum) observable is measured on both sides".

However, the latter claim is wholly misleading.

As proved in [2] in the frame of the quantum formalism, for a separable
quantum state, the validity of Eq. (\textrm{A6}) in [2] is not necessarily
linked with perfect correlations in this state. In view of the sufficient
conditions (53) derived in [2], there exist ideal quantum Alice/Bob joint
measurement situations, described by three quantum observables and performed
on a separable quantum state, where a separable state does not exhibit
perfect correlations but satisfies Eq. (\textrm{A6}) in [2]. What is most
important - there exist separable quantum states (cf. Eq. (49) in [2]) that
satisfy Eq. (\textrm{A6}) in [2] for any bounded quantum observables.

\emph{Concluding remarks.}\textit{\ }

In [2], the term "classical" is used according to its physical context.

The statement of [1] that the "definition of classicality" used in [2] "is
much narrower than Bell's concept of local hidden variables" constitutes the
misleading claim that the physical concept of classicality may depend on a
number of measured classical observables.

Under a classical Alice/Bob joint measurement of the same classical
observable on both sides, the outcomes are always perfectly correlated.
Therefore, the statement in [1] that "the author (of [2]) has implicitly
defined "classicality" to mean perfect correlations" is misleading.

The main statement in [1] is built up on arguments arising in the frame of a
Bell \textit{LHV} model for a separable quantum state and constitutes the
claim that, in the separable quantum case, the validity of the perfect
correlation form of the original Bell inequality\footnote{%
That is, Eq. (A6) in [2].} is necessarily linked with "the assumption of
perfect correlations if the same (quantum) observable is measured on both
sides". This claim contradicts the results derived in [2] in the frame of
the quantum formalism. Hence, this claim in [1] is misleading.

As proved in [2], there exist separable quantum states that satisfy the
perfect correlation form of the original Bell inequality for any bounded
quantum observables. These separable quantum states exhibit "classical-like"
character of behaviour under any ideal quantum Alice/Bob joint measurement
situations\footnote{%
Recall that a separable quantum state satisfies every CHSH-form inequality.}.

In conclusion, I would like to mention the new results in [4] where there is
introduced the whole class of quantum states, separable and nonseparable,
satisfying the perfect correlation form of the original Bell inequality
under any projective quantum measurements of Alice and Bob.

\end{document}